\documentclass[pra, aps, twocolumn,amssymb,amsmath,amsbsy, groupedaddress, superscriptaddress]{revtex4-1}
\usepackage{slashed}
\usepackage{graphicx}
\usepackage{float}
\usepackage{subcaption}
\usepackage{enumerate}
\usepackage{chngcntr}
\usepackage[usenames, dvipsnames]{color}
\usepackage{graphics,amsfonts,amsmath, amssymb}
\usepackage{mathtools}
\usepackage{hyperref}
\usepackage{bm}
\usepackage{amsfonts}
\usepackage{dsfont}
\usepackage{lipsum}
\usepackage[utf8]{inputenc}
\usepackage[T1]{fontenc}
\usepackage{bbold}
\usepackage{times}
\usepackage{color}

\usepackage[squaren,Gray]{SIunits}
\hypersetup{backref,
colorlinks=true,
urlcolor=blue,
linkcolor=blue,
linktoc=page,
citecolor=blue}

\everymath{\displaystyle} 

\begin{document}
	
	\title{Optimal Multiparameter Quantum Estimation of Magnonic Couplings in a Magnomechanical Cavity}
	\date{\today}
	\author{Adnan Naimy}
	\affiliation{LPHE, Modeling et Simulations, Faculty of Science, Mohammed V University in Rabat, Morocco.}
	\author{Abdallah Slaoui}
	\affiliation{LPHE, Modeling et Simulations, Faculty of Science, Mohammed V University in Rabat, Morocco.}
	\affiliation{CPM, Centre of Physics and Mathematics, Faculty of Science, Mohammed V University in Rabat, Morocco.}\affiliation{Center of Excellence in Quantum and Intelligent Computing, Prince Sultan University, Riyadh 11586, Saudi Arabia.}
	\author{Abderrahim Lakhfif}
	\affiliation{LPHE, Modeling et Simulations, Faculty of Science, Mohammed V University in Rabat, Morocco.}
	\affiliation{CPM, Centre of Physics and Mathematics, Faculty of Science, Mohammed V University in Rabat, Morocco.}
	\author{Rachid Ahl Laamara}
	\affiliation{LPHE, Modeling et Simulations, Faculty of Science, Mohammed V University in Rabat, Morocco.}
    \affiliation{CPM, Centre of Physics and Mathematics, Faculty of Science, Mohammed V University in Rabat, Morocco.}
\begin{abstract}
Enhanced quantum metrology plays a central role in revealing the fundamental features of quantum phenomena. In this context, the precise estimation of the magnon–cavity ($G_{mc}$) and magnon–mechanical ($G_{mb}$) coupling strengths is crucial for accurately characterizing and controlling the dynamics of a hybrid magnomechanical system. These two interactions govern the transfer of energy and information between the optical, magnetic, and mechanical subsystems, and optimizing their estimation significantly improves the systems performance in quantum metrology. In this work, we introduce an experimentally viable scheme to enhance the simultaneous estimation precision of the couplings $G_{mc}$ and $G_{mb}$, with a particular focus on the performance of heterodyne detection. By comparing simultaneous and individual estimation strategies, we demonstrate that the simultaneous approach offers a notable advantage in our system. To support this, we compute the quantum Fisher information matrices (QFIMs) based on the symmetric logarithmic derivative (SLD) and the right logarithmic derivative (RLD). Our results show that the quantum Cramér-Rao bound (QCRB) associated with the RLD is consistently lower than that of the SLD, indicating superior estimation precision. From a physical standpoint, this improvement reflects the system’s enhanced capacity to encode, transfer, and extract quantum information while allowing optimal control of fundamental interactions. We show that increasing the Rabi frequency, cavity loss rate, and the average number of photons and phonons, combined with reduced mechanical damping and temperature, enhances the system's sensitivity to the coupling parameters. These mechanisms act on the available quantum resources—such as entanglement, squeezing, and state purity—leading to more precise estimations. Furthermore, our analysis reveals that under certain conditions, heterodyne detection can closely approach the ultimate precision set by the QFIM. This suggests that a measurement strategy based on heterodyne detection can offer an efficient and practical route for estimating the couplings $G_{mc}$ and $G_{mb}$, paving the way for high-precision hybrid quantum sensors.
\end{abstract}

\maketitle
\section{INTRODUCTION}
Cavity optomechanics primarily investigates quantum phenomena arising from radiation pressure interactions between the optical and mechanical modes of the system \cite{Aspelmeyer2014}. In such systems, continuous-variable (CV) states—typically modeled as Gaussian states—encode and process quantum information via optical and mechanical modes \cite{Vitali2007,Manninen2018}. Over recent years, cavity optomechanical platforms have played a central role in the exploration of a variety of intriguing phenomena, including macroscopic quantum correlations \cite{Lai2022,Jiao2020,Vitali2007}, entangled quantum states \cite{Asjad2016,Lakhfif2024,Teklu2018}, mechanical oscillator cooling \cite{Frimmer2016,Huang2022}, photon and phonon blockade \cite{Rabl2011,Xie2017}, and optomechanically induced transparency \cite{Weis2010,Agarwal2010}. Moreover, these systems have demonstrated promising capabilities for enhancing precision measurements \cite{Liu2017,Xiong2017}. The ability to transfer quantum states between distinct subsystems is a crucial element for implementing protocols in quantum information processing and communication \cite{Asjad2016-2}.
In parallel, ferrimagnetic systems—particularly yttrium iron garnet (YIG) spheres—have garnered increasing attention within the framework of cavity quantum electrodynamics (QED). It has been shown that the Kittel mode \cite{Kittel1948} in a YIG sphere can strongly couple to microwave photons in high-quality-factor cavities, leading to the formation of cavity polaritons \cite{Huebl2013,Tabuchi2014,Zhang2014,Goryachev2014,Bai2015} and vacuum Rabi splitting. Several foundational concepts from cavity QED have thus been realized in this context, including optical bistability \cite{Wang2018}, as well as coherent interaction between a single superconducting qubit and the Kittel mode \cite{Tabuchi2015}.These advances emphasize the promising role of magnonic systems as an emerging platform for exploring strong-coupling regimes within cavity QED. Given this progress, it is therefore timely to investigate how magnons contribute to cavity optomechanics and the quantum phenomena they may enable. The first experimental realization of a magnon-photon-phonon interaction was recently reported \cite{Zhang2016}, demonstrating a setup in which photons couple to magnons—as in magnonic QED—and magnons, in turn, interact with phonons. The consequences of this hybrid interaction have been examined within the mean-field approximation, where quantum fluctuations are neglected, yet clear signatures of magnon-phonon interaction appear in the cavity output.\par
Building on these physical platforms, it becomes essential to explore the fundamental limits of parameter estimation and the role of quantum resources in optimizing measurement precision. Quantum parameter estimation theory aims to identify the ultimate bounds on the precision with which unknown quantities describing a quantum system can be measured, while simultaneously developing experimental strategies to approach or reach these limits. This field is of both technological relevance—especially in quantum-enhanced metrology and sensing \cite{Giovannetti2011,Demkowicz2015,Degen2017,Saidi2024,Pezze2018,Pirandola2018,Polino2020}—and foundational significance, being closely linked to concepts such as quantum state geometry \cite{Bengtsson2017} and the incompatibility of observables.\\
Early studies mainly focused on single-parameter estimation, leading to the design of protocols that exploit quantum resources to improve the sensitivity of physical measurements. In this context, the achievable precision is bounded by the quantum Cramér–Rao bound (QCRB), introduced by Helstrom \cite{Helstrom1969,Holevo2011,Braunstein1994,Paris2009}, under the assumption of repeated independent measurements. However, in many practical scenarios, it becomes necessary to estimate multiple parameters simultaneously \cite{Szczykulska2016,Demkowicz2020}, as in the case of orthogonal displacements \cite{Yuen1973,Hanamura2023}, waveform estimation \cite{Gardner2024}, multiphase interferometry \cite{Humphreys2013,Pezze2017,Abdellaoui2024,Gorecki2022}, joint estimation of phase and noise \cite{Pinel2013,Roccia2018,Abouelkhir2025}, and optomechanical systems \cite{Naimy2025-2}.\\
Additional applications include the localization of discrete optical sources \cite{Napoli2019,Bisketzi2019,Fiderer2021}, moment estimation of extended sources \cite{Tsang2019,Zhou2019}, and ranging and velocity measurements of moving targets \cite{Huang2021,Reichert2024}. Unlike the single-parameter case, the multiparameter scenario is complicated by the possible incompatibility of the optimal measurements associated with each parameter, which often prevents saturation of the QCRB. This challenge has motivated the development of quantum multiparameter metrology, which seeks to generalize the QCRB saturation conditions and design optimal estimation strategies.

At the core of this framework lies the Quantum Fisher Information Matrix (QFIM), whose inverse sets the ultimate lower bounds on the variances of simultaneously estimated parameters \cite{Braunstein1994,Naimy2025,Safranek2018}. Maximizing the QFIM is therefore a key objective in enhancing multiparameter estimation protocols. This matrix appears in various domains, including quantum thermometry \cite{Monras2011,Correa2015,Hofer2017}, squeezing parameter estimation \cite{Milburn1994,Chiribella2006}, and gravitational wave detection \cite{Abbott2016}. It also plays a central role in quantifying entanglement and coherence \cite{Hauke2016}, as well as in studying relativistic effects such as the Unruh–Hawking phenomenon \cite{Aspachs2010}.\\
To calculate the QFIM, especially in continuous-variable (CV) systems described by Gaussian states, various theoretical techniques have been developed. Gaussian states have garnered significant attention in quantum information science \cite{Ferraro2005,Braunstein2005}, primarily due to two key advantages: their tractability from an analytical standpoint, and their experimental accessibility. These states find applications in teleportation channels \cite{Wolf2007}, and cavity optomechanics \cite{Nunnenkamp2011,Peng2025}.

In this last context, radiation pressure interactions between optical and mechanical modes naturally give rise to Gaussian states \cite{Aspelmeyer2014}, which are involved in key phenomena such as mechanical cooling \cite{Frimmer2016,Huang2022} and optomechanically induced transparency \cite{Agarwal2010,Lai2020}. The relevance of the Gaussian formalism is further supported by experimental evidence in Bose–Einstein condensates \cite{Kevrekidis2003,Gross2011}. However, the complexity of the density matrix formalism makes the direct computation of the QFIM for Gaussian states challenging. To overcome this, many studies adopt a phase-space approach, where a Gaussian state is entirely described by its first and second moments. In this sense, integrating quantum multiparameter estimation theory with the analytical framework of Gaussian states becomes a fundamental step toward optimizing precision measurements in continuous-variable quantum systems. \par
In this work, we explore, from a theoretical perspective, the simultaneous estimation of two coupling parameters in an optomagnomechanical system consisting of a Fabry–Perot cavity embedded within a YIG (Yttrium Iron Garnet) sphere. A microwave field is applied to enhance the magnon–phonon coupling. At the location of the YIG sphere, three orthogonal magnetic fields are applied: a static bias field along the $z$-axis, the magnetic field of the cavity mode along the $x$-axis, and a driving field along the $y$-axis.. More specifically, we consider a hybrid cavity–magnon–mechanical system \cite{Zhang2016} comprising photons, magnons (collective spin excitations in a ferromagnetic medium), and phonons. As described in Ref. \cite{Zhang2016}, a YIG sphere with a diameter of \unit{250}{\micro\meter} is used. In this setup, the magnon–photon coupling is mediated by magnetic dipole interaction, while the magnon–phonon interaction is governed by magnetostrictive effects. In particular, magnon excitations within the YIG sphere modify its local magnetization, which leads to deformations of its geometric structure, thereby generating vibrational modes (phonons), and vice versa \cite{Kittel1958}.\par 

To assess the ultimate precision in parameter estimation, we analyze the QFIM using both the Right Logarithmic Derivative (RLD) \cite{Holevo2003} and the Symmetric Logarithmic Derivative (SLD) \cite{Safranek2018}. The phase-space formalism proves particularly advantageous in this context, as it greatly simplifies the computation of key quantities in multiparameter quantum estimation and yields analytically tractable results for continuous-variable systems.\par

This article is structured to guide the reader progressively through the different stages of our analysis. In Sec. \ref{sec_2}, we introduce the cavity magnomechanical system under study, describing the physical model as well as the resulting dynamical equations. Sec. \ref{sec_3} is dedicated to the theoretical foundations of multiparameter quantum estimation, particularly in the context of Gaussian states. We derive analytical expressions for the QFIM based on the SLD and RLD operators, and compare the performance of the individual and simultaneous estimation strategies using a quantitative ratio. This section also presents the calculation of the classical Fisher information (CFI) using the heterodyne detection method. In Sec. \ref{sec_4}, the theoretical tools developed earlier are applied to the system under consideration to derive explicit expressions for the QFIM-SLD, QFIM-RLD, and CFI, highlighting their practical implications. Finally, Sec. \ref{sec_5} concludes this work with a summary of the main results obtained.
\section{Theoretical Model and Dynamical Analysis} \label{sec_2}

\begin{figure}[H]
    \centering
    \includegraphics[width=0.44\textwidth]{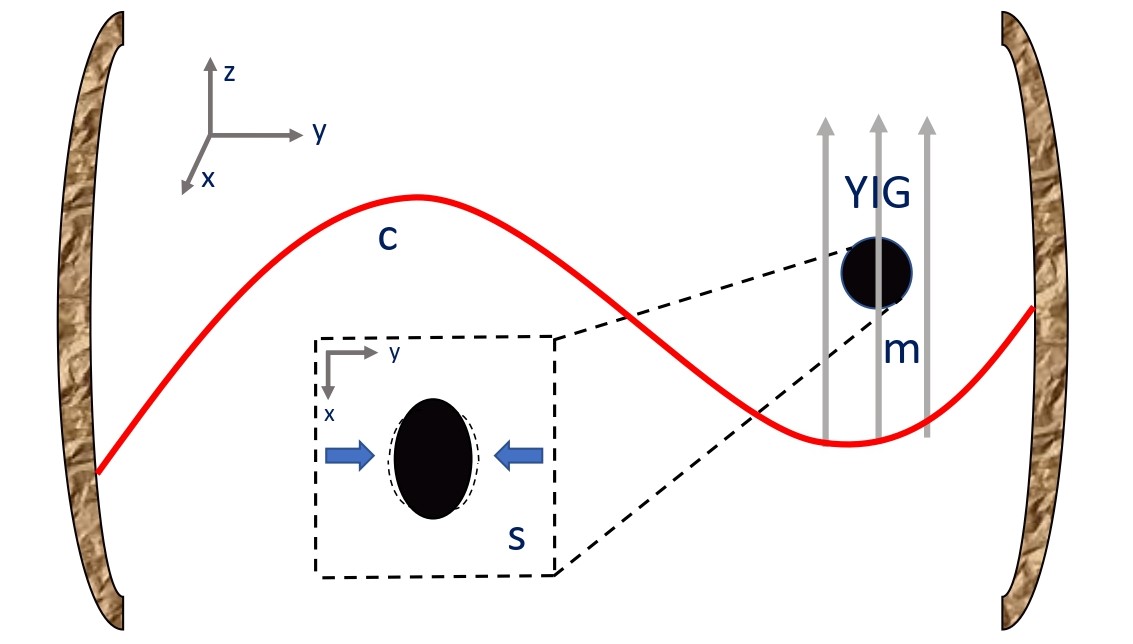}
    \caption{The schematic illustrates a hybrid magnomechanical system where a YIG sphere is placed inside a microwave cavity at the point of maximum magnetic field intensity of the cavity mode. A uniform static magnetic field is applied to establish magnon–photon interaction. To enhance the magnomechanical coupling, the magnon mode is externally driven by a microwave source (not depicted). At the sphere’s location, three mutually orthogonal magnetic fields are present: the static bias field (along the $z$-axis), the external drive field (along the $y$-axis), and the magnetic field of the cavity mode (along the $x$-axis).} 
    \label{schema}
\end{figure}

As illustrated in Fig.\ref{schema}, we consider a hybrid cavity magnomechanical system operating in the microwave domain, consisting of interacting photons, magnons, and phonons \cite{Zhang2016}. Magnons represent collective spin excitations within a ferromagnetic medium. A typical implementation, such as in Ref. \cite{Zhang2016}, involves a yttrium iron garnet (YIG) sphere with a diameter of \unit{250}{\micro\meter}, which is small enough relative to the microwave wavelength to neglect radiation pressure effects. The magnon–photon interaction arises via magnetic dipole coupling, whereas the magnon–phonon interaction is facilitated through magnetostrictive effects. More specifically, the excitation of magnons inside the YIG sphere induces a variation in magnetization, which leads to a deformation of the sphere’s geometric structure, thereby generating vibrational modes (phonons), and vice versa \cite{Kittel1958}. 

The total Hamiltonian of the system reads
\begin{align}
    \mathcal{H}/\hbar=& \, \omega_{c} c^+c + \omega_{m} m^+m + \frac{\omega_s}{2}(q^2+p^2) + g_{ms}m^+mq \nonumber\\
    &+G_{mc} (c+c^+)(m+m^+) \nonumber \\
    &+i  \Omega \big(m^+e^{-i\omega_d \textit{t}}-me^{i\omega_d \textit{t}} \big). \label{Hamiltonien}
\end{align}
The first two terms of the Hamiltonian correspond to the free energy of the optical and magnonic modes, described by the bosonic operators \( c^\dagger, c \) and \( m^\dagger, m \), which obey the commutation relations \( [\mathcal{O}, \mathcal{O}^\dagger] = 1 \), where \( \mathcal{O} = c, m \). The third term represents the mechanical mode, described in terms of the dimensionless position and momentum quadratures \( q \) and \( p \), satisfying \( [q, p] = i \). The resonance frequencies of the cavity, magnon, and mechanical modes are denoted by \( \omega_c \), \( \omega_m \), and \( \omega_s \), respectively, with the magnon frequency given by \( \omega_m = \gamma H \), where \( \gamma \) is the gyromagnetic ratio and \( H \) is the external magnetic bias field. The fourth term characterizes the magnetostrictive interaction between the magnon and mechanical modes, with a single-magnon–phonon coupling rate \( g_{ms} \), which is typically weak but can be enhanced via strong microwave driving of the magnon mode. Such excitation is generally implemented by directly driving the YIG sphere with a microwave field~\cite{Wang2018,Wang2016}. The fifth term corresponds to the magnon–photon interaction governed by the coupling rate \( G_{mc} \), arising from magnetic dipole interaction. When \( G_{mc} > \kappa_c, \kappa_m \), the system enters the strong coupling regime~\cite{Huebl2013,Tabuchi2014,Zhang2014,Goryachev2014,Bai2015}. Finally, the last term accounts for the coherent drive applied to the magnon mode by an external microwave source, intended to reinforce the effective magnomechanical coupling.\par

The corresponding Rabi frequency, given by $\Omega = \frac{5}{4} \gamma \sqrt{N B_0}$, characterizes the coupling strength between the oscillating magnetic field—of frequency $\omega_d$ and amplitude $B_0$—and the magnon mode in the YIG sphere. Here, $\gamma/2\pi = \unit{28}{GHz/T}$, and the total number of spins is $N = \rho\,V$, where $\rho = 4.22 \times 10^{27}\, \mathrm{m^{-3}}$ is the spin density of YIG and $V$ is the volume of the sphere. To enhance the interaction, the YIG sphere is placed at a location of maximum magnetic field intensity within the cavity. The expression for $\Omega$ assumes the low-excitation regime, i.e., $\langle m^\dagger m \rangle \ll 2NS$, where $S = 5/2$ corresponds to the spin of $\mathrm{Fe}^{3+}$ ions in YIG.\\
In the rotating frame defined by the drive frequency $\omega_0$, and assuming the regime $\omega_m, \omega_c \gg G_{mc}, \kappa_m, \kappa_c$, one can invoke the rotating-wave approximation (RWA). Under this approximation, the interaction term $G_{mc} (c + c^\dagger)(m + m^\dagger)$ is simplified to $G_{mc} (c\,m^\dagger + c^\dagger\,m)$, which neglects fast-oscillating counter-rotating terms. This condition is well justified in the experimental parameters reported in Ref.~\cite{Zhang2016}. The dynamics of the system are described by the following quantum Langevin equations ({\bf QLEs})
\begin{subequations}
    \begin{align}
        \partial_t\, c =& -(i\Delta_c+\kappa_c)c-iG_{mc}\,m+\sqrt{2\kappa_c} \,c^{in},\\
        \partial_t\,m =& -iG_{mc}\,c -(i\Delta_m+\kappa_m)m - ig_{ms}\,mq  \nonumber\\ 
        &+ \Omega+\sqrt{2\kappa_m} \,m^{in},\\
        \partial_t\,q=& \;\omega_s\,p,\\
        \partial_t\,p=& - g_{ms} \,m^\dagger m -\omega_s\,q-\gamma_s\,p  + \eta .
    \end{align}
\end{subequations}
The detunings of the cavity and magnon modes from the driving frequency are defined as $\Delta_c = \omega_c - \omega_d$ and $\Delta_m = \omega_m - \omega_d$, respectively. The mechanical mode is subject to a damping rate denoted by $\gamma_s$. The input noise operators $c^{\text{in}}$, $m^{\text{in}}$, and $\eta$ account for the quantum fluctuations entering the cavity, magnon, and mechanical modes. These operators have zero mean and satisfy the standard correlation functions provided in Ref.~\cite{Li2018}
\begin{subequations}
    \begin{align}
        &\langle c^{in}(\textit{t})\; c^{in,\dagger}(\textit{t}')\rangle =\left[ \textit{N}_c(\omega_c)+1\right] \,\delta(\textit{t}-\textit{t}'), \\
        &\langle c^{in,\dagger}(\textit{t})\; c^{in}(\textit{t}')\rangle = \textit{N}_c(\omega_c) \,\delta(\textit{t}-\textit{t}'), \\
        &\langle m^{in}(\textit{t})\; m^{in,\dagger}(\textit{t}')\rangle = \left[\textit{N}_m(\omega_m)+1\right] \,\delta(\textit{t}-\textit{t}'), \\
        &\langle m^{in,\dagger}(\textit{t})\; m^{in}(\textit{t}')\rangle = \textit{N}_m(\omega_m) \,\delta(\textit{t}-\textit{t}'), \\
        &\langle \eta(\textit{t})\eta(\textit{t}')+ \eta(\textit{t}')\eta(\textit{t})\rangle =2\,\gamma_s\left[2 \text{N}_s(\omega_s)+1\right] \,\delta(\textit{t}-\textit{t}').
    \end{align}
\end{subequations}
Here, $\textit{N}_j(\omega_j)=\left(\text{exp}[\hbar \omega_j/K_BT]-1\right)^{-1}$ (with $j=c,m,s$) denotes the mean equilibrium thermal occupation number of photons, magnons, and phonons, respectively. Moreover, under the assumption that the mechanical resonator operates in the high-quality-factor regime, i.e., $Q=\omega_s/\gamma_s \gg1$, a Markovian approximation has been employed, which is well justified in this regime \cite{Benguria1981,Giovannetti2001}.\par
In the regime where the magnon mode is subject to strong external driving, it acquires a significant steady-state amplitude, such that $|\langle m \rangle| \gg 1$. Owing to the beam-splitter-type interaction between the magnon and cavity modes, the cavity field also reaches a large steady-state value, satisfying $|\langle c \rangle| \gg 1$. These conditions allow us to linearize the system's dynamics by decomposing each operator into its steady-state part and small quantum fluctuations, i.e., $\mathcal{O} = \mathcal{O}_a + \delta\mathcal{O}$, with $\mathcal{O} \in \{c, m, q, p\}$. Second-order fluctuation terms are neglected in this approximation. Under this linearization, the dynamics of the fluctuation operators can be described by a set of QLEs for the quadrature variables $\delta X_c, \delta Y_c, \delta X_m, \delta Y_m, \delta q, \delta p$. These are defined as $\delta X_c = (\delta c^\dagger + \delta c)/\sqrt{2}$, $\delta Y_c = i(\delta c^\dagger - \delta c)/\sqrt{2}$, $\delta X_m = (\delta m^\dagger + \delta m)/\sqrt{2}$, and $\delta Y_m = i(\delta m^\dagger - \delta m)/\sqrt{2}$. The corresponding linearized QLEs are given by
\begin{equation}
    \dot{\text{u}}(\textit{t}) = \mathcal{A}\, \text{u}(\textit{t})+\text{n}(\textit{t}) ,
\end{equation}
where, $\text{n}(\textit{t})=\big[\sqrt{2\kappa_c} \,X_c^{in}(\textit{t}), \sqrt{2\kappa_c} \,Y_c^{in}(\textit{t}),\sqrt{2\kappa_m} \,X_m^{in}(\textit{t}),$\\$ \sqrt{2\kappa_m} \,Y_m^{in}(\textit{t}), 0, \eta(\textit{t})\big]^T$ represents the input noise vector, $\text{u}(\textit{t})=\left[\delta X_c(\textit{t}), \delta Y_c(\textit{t}), \delta X_m(\textit{t}), \delta Y_m(\textit{t}), 0, \eta(\textit{t})\right]^T$ denotes the vector of quadrature fluctuations, and the systems dynamics are governed by the drift matrix $\mathcal{A}$, which can be expressed as follows
\begin{equation}
	\mathcal{A}=\left(\begin{array}{cccccc}
               -\kappa_c & \Delta_c & 0 & G_{mc} & 0 & 0  \\
			 -\Delta_c & -\kappa_c & -G_{mc} & 0 & 0 & 0 \\
			 0 & G_{mc} & -\kappa_m & \Delta'_m & -G_{ms} & 0  \\
			 -G_{mc} & 0 & -\Delta'_m & -\kappa_m & 0 & 0 \\
             0 & 0 & 0 & 0 & 0 & \omega_s \\
             0 & 0 & 0 & G_{ms} & -\omega_s & -\gamma_s 
             
	\end{array}\right). \label{matrix A}
\end{equation}
Here, $G_{ms}=i\sqrt{2} g_{ms}\, m_a$ denotes the effective magnomechanical coupling rate, and $\Delta'_m =\Delta_m+g_{ms}\,q_a$ is the effective magnon-drive detuning, which includes the shift in frequency resulting from the magnomechanical interaction. The mean displacement $q_a$ is given by
\begin{equation}
    q_a= - \frac{g_{ms}}{\omega_s} |m_a|^2,
\end{equation}
where the magnon mode attains the following steady-state amplitude
\begin{equation}
    m_a=\frac{\Omega(i\Delta_c+\kappa_c)}{G_{mc}^2+(i\Delta_c+\kappa_c)\;(i\Delta'_m+\kappa_m)} .\label{ms}
\end{equation}
Equation~\eqref{matrix A} assumes a regime where the detunings $|\Delta_c|$ and $|\Delta'_m|$ are much larger than their associated damping rates $\kappa_c$ and $\kappa_m$, respectively. As will be demonstrated later, the regime $|\Delta_c|, |\Delta'_m| = \omega_s \gg \kappa_c, \kappa_m$  proves to be optimal for enhancing the precision of multiparameter estimation within the hybrid system. It is important to point out that Eq. \eqref{ms} is fundamentally nonlinear due to the dependence of $|\Delta'_m|$ on $|m_a|^2$ . Nevertheless, for any chosen value of $\Delta_m$—which can be tuned via the external bias magnetic field—the corresponding steady-state amplitude $m_a$, and hence the effective coupling $G_{ms}$, can be readily determined.\par

Due to the system's linearized dynamics and the Gaussian nature of its noise sources, the quantum fluctuations evolve towards a stationary state that corresponds to a continuous-variable (CV) Gaussian state involving three interacting modes. This state can be entirely described by a $6\times6$ covariance matrix $\mathcal{V}$, whose components are defined as $\mathcal{V}_{ij} = \frac{1}{2} \langle u_i(\textit{t}), u_j(\textit{t}') + u_j(\textit{t}'), u_i(\textit{t}) \rangle$ for $i,j = 1,\dots,6$. The explicit form of the steady-state covariance matrix is obtained by solving the corresponding Lyapunov equation \cite{Vitali2007, Parks11993}
\begin{equation}
    \mathcal{A} \mathcal{V} + \mathcal{V} \mathcal{A}^T=-\mathcal{D},
\end{equation}
where $\mathcal{D}=\text{diag} \big[\kappa_c (2 \textit{N}_c+1), \kappa_c(2 \textit{N}_c+1), \kappa_m(2 \textit{N}_m+1), \kappa_m(2 \textit{N}_m+1), 0, \gamma_s(2 \textit{N}_s+1)\big]$
represents the diffusion matrix, which characterizes $\langle \text{n}_i(\textit{t})\,\text{n}_j(\textit{t}')+\text{n}_j(\textit{t}')\,\text{n}_i(\textit{t})\rangle=2\, \mathcal{D}_{ij}\delta(\textit{t}-\textit{t}')$.
\section{Classical and Quantum Estimation Theory in Gaussian States} \label{sec_3}

In both classical and quantum frameworks, a physical system can be fully described either by a covariance matrix \( \mathcal{V} \) or by a positive semi-definite density operator \( \hat{\rho}(\boldsymbol{\varphi}) \), where \( \boldsymbol{\varphi} = \{ \varphi_1, \varphi_2, ..., \varphi_n \} \) denotes a set of unknown parameters to be estimated. These parameters are typically encoded into the system’s dynamics through an evolution map \( \theta_{\boldsymbol{\varphi}} \), which acts on a probe state \( \rho \) to produce the parameter-dependent output state \( \rho_{\boldsymbol{\varphi}} \). A generalized quantum measurement described by a POVM \( \{ \Pi_x \} \) is then performed, producing outcomes with probabilities \( P(x|\boldsymbol{\varphi}) = \mathrm{Tr}[\rho_{\boldsymbol{\varphi}} \Pi_x] \). From these outcomes, an estimator \( \hat{\boldsymbol{\varphi}} \) is constructed to infer the values of the parameters. Quantum parameter estimation theory aims to identify the fundamental limits on how precisely unknown parameters can be estimated, typically expressed through the QCRB. This bound sets a fundamental limit on the covariance of any unbiased estimator, dictated by the QFIM. The two most widely used formulations of the QFIM rely on the SLD and RLD, introduced through the following differential relations~\cite{Genoni2013, Gao2014, Fujiwara1994} 
\begin{align}
     \partial_{\varphi_\mu} \hat{\rho} =& \,\hat{\rho}\; \hat{\mathbf{L}}^R_{\varphi_\mu},\hspace{0.5cm}(RLD), \label{RLD generale}\\
    \partial_{\varphi_\mu} \hat{\rho} =&\, \frac{1}{2} \{\hat{\rho},\hat{\mathbf{L}}^S_{\varphi_\mu}\},\, \hspace{0.5cm}(SLD), \label{SLD generale}
\end{align}
where $\hat{\mathbf{L}}^R_{\varphi_\mu}$ and $\hat{\mathbf{L}}^S_{\varphi_\mu}$ represent the RLD and SLD operators, respectively. $\{.,.\}$ denotes the anticommutator between two operators, and $\partial_{\varphi_\mu}$ indicates a partial derivative. Subsequently, we can establish two distinct matrices \cite{Gao2014}
\begin{align}
      \mathrm{F}_{\varphi_\mu \varphi_\nu} =& Tr\big[\hat{\rho} \hat{\mathbf{L}}^{R}_{\varphi_\mu}\, \hat{\mathbf{L}}^{{R}\dagger}_{\varphi_\nu}\big],\label{H RLD}\\
     \mathrm{H}_{\varphi_\mu\varphi_\nu} =& \, \frac{1}{2} Tr\big[\hat{\rho}\{\hat{\mathbf{L}}^S_{\varphi_\mu},\; \hat{\mathbf{L}}^S_{\varphi_\nu}\}\big]. \label{F SLD}
\end{align}
These are referred to as the RLD-QFIM and SLD-QFIM, respectively. The QFIM-SLD can be derived from Uhlmann’s quantum fidelity between two final output states corresponding to different parameter sets \cite{Braunstein1994}. The two distinct forms of the QCRB can be established \cite{Gao2014,Helstrom1976}
\begin{align}
    \text{ Cov}(\hat{\varphi}) &\ge \frac{1}{\mathcal{N}}\,\left(\mathrm{F}^{-1}_R+\big|\mathrm{F}^{-1}_I\big|\right), \label{covH}\\
     \text{ Cov}(\hat{\varphi}) &\ge \frac{1}{\mathcal{N}}\, \mathrm{H}^{-1}. \label{covF}
 \end{align}
Let 
$\mathrm{F}^{-1}=\,\mathrm{F}_R^{-1}+i\mathrm{F}_I^{-1}$ , where $\mathrm{F}_R$ and $\mathrm{F}_I$ refer to the real and imaginary components of the matrix $\mathrm{F}$, respectively. The modulus of any quantity '$\text{v}$' is given by $|\text{v}|=\sqrt{\text{v}\:\text{v}^\dagger}$. In this context, $\mathcal{N}$ designates the total number of repeated measurements, and $\text{Cov}(\hat{\varphi})$ represents the covariance matrix, defined as follows \cite{Gordon2022}
\begin{align}
     \text{Cov}(\varphi_\mu,\,\varphi_\nu) =& \,E \big[[\varphi_\mu-E(\varphi_\mu)]\,[\varphi_\nu-E(\varphi_\nu)] \big] \nonumber \\
     =&\, E(\varphi_\mu \varphi_\nu)- E(\varphi_\mu) E(\varphi_\nu). 
 \end{align}
Specifically, the individual estimation strategy corresponds to the condition $\mathrm{F}_{\varphi_\mu \varphi_\nu}=\mathrm{F}_{\varphi_\mu \varphi_\nu}=0$ when $\mu \neq \nu$. Under this condition, the precision in estimating a parameter is optimally characterized by its variance, leading to a simplified form of Eqs. \eqref{covH} and \eqref{covF}
\begin{align}
    \text{Var}(\varphi_\mu) &\ge \frac{1}{\mathcal{N}}\,\big( \mathrm{F}^{-1}_R+\big|\mathrm{F}^{-1}_{I}\big|\big), \label{varF}\\
     \text{Var}(\varphi_\mu) &\ge \frac{1}{\mathcal{N}}\, \mathrm{H}^{-1}.\label{varH}
\end{align}
Eqs.~\eqref{varF} and \eqref{varH} are always saturated, indicating that the corresponding parameter estimation achieves optimal precision. This saturation is attained when the measurement is performed in the eigenbasis of the SLD, with the optimal states given by the projectors onto its eigenvectors. By applying the trace operator to inequalities \eqref{covH} and \eqref{covF}, one obtains expressions corresponding to the overall variance across all estimated parameters
\begin{align}
\sum_\mu^N \text{Var}(\varphi_\mu) \ge&\; \mathcal{B}_R = \,\frac{1}{\mathcal{N}}\,\big( Tr\big[\mathrm{F}^{-1}_{R}\big]+Tr\big[\big|\mathrm{F}^{-1}_{I}\big|\big]\big),\\
\sum_\mu^N \text{Var}(\varphi_\mu) \ge&\; \mathcal{B}_S = \, \frac{1}{\mathcal{N}}\, Tr\big[\mathrm{H}^{-1}\big].
\end{align}
Typically, neither the RLD-bound nor the SLD-bound can be exactly reached \cite{Monras2011}.This limitation stems from the intrinsic non-commutative nature of quantum mechanics, which hinders the ability to achieve optimal precision when estimating multiple parameters simultaneously. In practice, enhancing the precision for one parameter often degrades the accuracy for others. It is important to note that the estimator achieving the RLD-bound is not always linked to a physically implementable POVM. However, in scenarios where the optimal measurements for individual parameters are non-commuting, it remains possible to attain both RLD and SLD bounds by employing an appropriately engineered joint measurement strategy. A substantial portion of the research in quantum parameter estimation has predominantly focused on the SLD-based bound, as explored in works such as \cite{Monras2011,Monras2010,Crowley2014}, with further developments presented in \cite{Helstrom1974,Genoni2013}. In single-parameter estimation, it is established that the SLD QFI is always less than or equal to the RLD QFI, thus providing a stricter bound on sensitivity \cite{Helstrom1969,Monras2011}. In the context of multiple parameters, recent advancements in the theory of quantum local asymptotic normality demonstrate that Eq. \eqref{varH} is asymptotically achievable under the condition
\begin{equation}
Tr\left[\hat{\rho}\left[\hat{\mathbf{L}}^S_{\varphi_\mu}, \hat{\mathbf{L}}^S_{\varphi_\nu}\right]\right] = 0. \label{cond rho Ls}
\end{equation}
According to Ref.~\cite{Nichols2018}, the above condition can be equivalently expressed as
\begin{equation}
    \text{Im} \left(Tr\left[\hat{\rho}\, \hat{\mathbf{L}}^{S}_{\varphi_\mu}\, \hat{\mathbf{L}}^{S}_{\varphi_\nu}\right]\right)=0.
\end{equation}
When the required condition is not satisfied, the RLD bound may offer a tighter limit and thus becomes particularly relevant. Deriving analytical forms of the RLD and SLD operators, as well as the corresponding QFIMs, remains a challenging task—except in special cases where the density matrix can be easily diagonalized \cite{Monras2007,Aspachs2010,Zhang2013}. For Gaussian states, however, the authors of Ref. \cite{Monras2013} successfully derived explicit expressions for both the SLD and RLD operators, along with their associated QFIMs. In this work, we apply these results to a magnonic system in order to compute the corresponding QCRBs. This naturally leads to the question; How are the RLD-based bound $\mathcal{B}_R$ and the SLD-based bound $\mathcal{B}_S$ related ? and which of these provides more informative insight ? these questions have been addressed in Refs. \cite{Genoni2013,Gao2014} by introducing the concept of the Most Informative-QCRB ($ \mathcal{B}_{MI}$), defined as
\begin{equation}
    \mathcal{B}_{MI} = \text{Min} \{\mathcal{B}_R,\mathcal{B}_S \}.\label{BMI}
\end{equation}
To identify which of the two bounds, $\mathcal{B}_R$ or $\mathcal{B}_S$, is smaller in Eq. \eqref{BMI}, we evaluate their ratio by directly comparing them, as follows
\begin{equation}
    \mathcal{R} = \frac{\mathcal{B}_R}{\mathcal{B}_S}.\label{R genarale}
\end{equation}
In general, the ratio $\mathcal{R}$ falls into one of three distinct categories, each reflecting a different comparative behavior between the RLD and the SLD bounds. These cases provide insight into which bound offers tighter precision depending on the characteristics of the system. Specifically, if $\mathcal{R}>1$ , the most informative bound is $\mathcal{B}_S$. If $\mathcal{R}<1$, the most informative bound is $\mathcal{B}_R$. Finally, when $\mathcal{R}=1$, both bounds yield the same precision, that is, $\mathcal{B}_S=\mathcal{B}_R$.\\
As shown in Eq. \eqref{BMI}, the optimal measurements in multiparameter estimation protocols are determined exclusively by the most informative QCRB, which is characterized by the following inequality
\begin{equation}
    \sum_\mu^N \text{Var}(\varphi_\mu) \ge \, \frac{1}{\mathcal{N}}\, \mathcal{B}_{MI}.
\end{equation}
Within the framework of quantum metrology, the QFI provides a more precise parameter estimation than the classical Fisher Information (CFI). In this section, we explicitly derive the SLD and the RLD operators for quantum Gaussian states. Based on these results, we compute the corresponding QFIMs associated with each operator. Finally, we present the general analytical expression of the CFI derived from the SLD operator in the context of single-parameter estimation.

\subsection{Quantum context}

In this section, we present the explicit derivation of both the SLD and the RLD operators within the formalism of quantum Gaussian states. These operators are essential for evaluating the QFI. Utilizing their analytical forms, we subsequently determine the corresponding expressions for the QFIM based on the SLD and RLD approaches. For an N-mode Gaussian state, we assume that the operators $\hat{\mathbf{L}}^R_{\varphi_\mu}$ and $\hat{\mathbf{L}}^S_{\varphi_\mu}$, defined by Eqs. \eqref{RLD generale} and \eqref{SLD generale}, can be represented as quadratic forms in the canonical operators. This assumption allows us to express the corresponding Fisher information matrices in a compact and analytical form, as given by Eq. \eqref{H RLD} for the RLD, and Eq. \eqref{F SLD} for the SLD. For clarity, we adopt the Einstein summation convention over repeated indices throughout this work
\begin{align}
     \hat{\mathbf{L}}^R_{\varphi_\mu}=&\; \hat{\mathbf{L}}^{R_0} + \hat{\mathbf{L}}^{R_1}_j \hat{{\text{r}}}_j + \hat{\mathbf{L}}^{R_2}_{k l} \hat{{\text{r}}}_k \hat{{\text{r}}}_l, \hspace{1cm}\text{(RLD)}\\
      \hat{\mathbf{L}}^S_{\varphi_\mu}=& \;\hat{\mathbf{L}}^{S_0} + \hat{\mathbf{L}}^{S_1}_j \hat{{\text{r}}}_j + \hat{\mathbf{L}}^{S_2}_{k l} \hat{{\text{r}}}_k \hat{{\text{r}}}_l, \hspace{1.15cm}\text{(SLD)}
\end{align}
where $\hat{{\text{r}}} = ( \hat{\text{q}}_1, \hat{\text{p}}_1, ..., \hat{\text{q}}_n, \hat{\text{p}}_n )^\mathbf{T}$ denotes the transposed vector of canonical operators, and the expressions of the components involved in the definition of the SLD and RLD operators are presented below; the detailed calculations can be found in Refs. \cite{Safranek2018,Gao2014}
\begin{subequations}
\begin{align}
  &\hat{\mathbf{L}}^{R_0}_{\varphi_\mu} = -\frac{1}{2}Tr\left[\mathcal{J}_+  \hat{\mathbf{L}}^{R_2}_{\varphi_\mu} \right] - d_r^\mathbf{T}  \hat{\mathbf{L}}^{R_1}_{\varphi_\mu} - d_r^\mathbf{T}  \hat{\mathbf{L}}^{R_2}_{\varphi_\mu}d_r\in  \mathbb{R},\\
  &\hat{\mathbf{L}}^{R_1}_{\varphi_\mu} = 2 \mathcal{J}_+^{-1} \partial_{\varphi_\mu} d_r - 2 \hat{\mathbf{L}}^{R_2}_{\varphi_\mu}d_r\, \in \mathbb{R}_{2N},\\
  &\text{vec} \left[ \hat{\mathbf{L}}^{R_2}_{\varphi_\mu} \right] = \left(\mathcal{J}^\dagger \otimes \mathcal{J} \right)^+ \text{vec} \left[\partial_{\varphi_\mu} \mathcal{V} \right]\;\in \mathbb{R}_{2N \times 2N}, 
\end{align} 
\end{subequations}
and 
\begin{subequations}
\begin{align}
 &\hat{\mathbf{L}}^{S_0}_{\varphi_\mu} =-\frac{1}{2}Tr\left[\mathcal{V}\;  \hat{\mathbf{L}}^{S_2}_{\varphi_\mu} \right] - d_r^\mathbf{T}  \hat{\mathbf{L}}^{S_1}_{\varphi_\mu} - d_r^\mathbf{T}  \hat{\mathbf{L}}^{S_2}_{\varphi_\mu}d_r \,\in  \mathbb{C}, \\
      &\hat{\mathbf{L}}^{S_1}_{\varphi_\mu} = 2V^{-1} \partial_{\varphi_\mu} d_r - 2 \hat{\mathbf{L}}^{S_2}_{\varphi_\mu}d_r\, \in \mathbb{C}_{2N},\\
      &\text{vec} \left[ \hat{\mathbf{L}}^{S_2}_{\varphi_\mu} \right] = \left(\mathcal{V}^\dagger \otimes \mathcal{V} + \omega \otimes \omega \right)^+ \text{vec} \left[\partial_{\varphi_\mu} \mathcal{V} \right] \, \in \mathbb{C}_{2N \times 2N} .
\end{align} 
\end{subequations}
Let $\mathcal{J}= 2\mathcal{V}+i\Omega$ be a $2N\times2N$ matrix, where $\mathcal{V}$ is the covariance matrix, and 
$\omega=\oplus_{k=1}^n i\,\sigma_y$, with $\sigma_y$ being a Pauli matrix. Here, $\text{vec}[\bullet]$ denotes the vectorization of any $n\times n$ complex or real matrix, defined as $$\text{vec}[\bullet]=[\text{v}_{11}, ..., \text{v}_{n1}, \text{v}_{12}, ..., \text{v}_{n2}, \text{v}_{1n}, ..., \text{v}_{nn}]^\mathbf{T}.$$
 By inserting the expressions of the RLD and SLD operators into Eqs. \eqref{H RLD} and \eqref{F SLD}, we obtain the QFIM-RLD and the QFIM-SLD, respectively (see Refs. \cite{Safranek2018,Gao2014})
\begin{equation}
     \mathrm{F}_{\varphi_\mu\varphi_\nu} = 2\, \text{vec}\left[\partial_{\varphi_\mu}\mathcal{V} \right]^\dagger  
     \mathrm{\Sigma}^{+}  \text{vec}\left[\partial_{\varphi_\nu}\mathcal{V} \right] + 2 \partial_{\varphi_\mu}  d_r^{\mathbf{T}} \mathcal{J}^{+} \partial_{\varphi_\nu}  d_r,\label{H with inv}
\end{equation}
\begin{equation}
     \mathrm{H}_{\varphi_{\mu} \varphi_{\nu}} = \, 2\,\text{vec}\left[\partial_{\varphi_\mu}\mathcal{V} \right]^\dagger  
     \mathfrak{M}^{+}  \text{vec}\left[\partial_{\varphi_\nu}\mathcal{V} \right] +  \partial_{\varphi_\mu}  d_r^{\mathbf{T}} \mathcal{V}^{-1} \partial_{\varphi_\nu}  d_r.\label{F with inv}
\end{equation}
Let us define $\mathrm{\Sigma}^+=(\mathcal{J}^\dagger\otimes\mathcal{J})^+$ and 
$\mathfrak{M}^+= \left(4\,\mathcal{V}^\dagger \otimes\mathcal{V}+\Omega\otimes\Omega\right)^+$,
where the superscript '$+$' denotes the Moore–Penrose pseudoinverse, a generalization of the standard matrix inverse \cite{Penrose1955,Ben-Israel2006}. This pseudoinverse can be computed, for instance, using Tikhonov regularization \cite{Golub2013} $\text{A}^+=\lim\limits_{x \rightarrow 0} \left(\text{A}^\dagger(\text{A}\text{A}^\dagger+ \delta I)^{-1}\right)=\lim\limits_{x \rightarrow 0} \left((\text{A}\text{A}^\dagger+ \delta I)^{-1}\text{A}^\dagger\right)$, which ensures the existence of a stable approximation even in cases where $\text{A}^{-1}$ does not exist. Furthermore, when both $\mathcal{J}$ and $\mathfrak{M}$ are invertible (i.e., non-singular), the expressions of the QFIM-RLD and the QFIM-SLD simplify significantly and can be written in closed form
\begin{equation}
     \mathrm{F}_{\varphi_\mu\varphi_\nu} = 2\, \text{vec}\left[\partial_{\varphi_\mu}\mathcal{V} \right]^\dagger  
     \mathrm{\Sigma}^{-1}  \text{vec}\left[\partial_{\varphi_\nu}\mathcal{V} \right] + 2 \partial_{\varphi_\mu}  d_r^{\mathbf{T}} \mathcal{J}^{-1} \partial_{\varphi_\nu}  d_r,\label{F with inv}
\end{equation}
\begin{equation}
     \mathrm{H}_{\varphi_{\mu} \varphi_{\nu}} = \, 2\,\text{vec}\left[\partial_{\varphi_\mu}\mathcal{V} \right]^\dagger  
     \mathfrak{M}^{-1}  \text{vec}\left[\partial_{\varphi_\nu}\mathcal{V} \right] +  \partial_{\varphi_\mu}  d_r^{\mathbf{T}} \mathcal{V}^{-1} \partial_{\varphi_\nu}  d_r.\label{H with inv}
\end{equation}
\par
We now consider two widely adopted strategies in multiparameter quantum metrology, both recognized for their effectiveness in extracting information about unknown parameters. The individual estimation strategy involves estimating each parameter independently, under the assumption that all other parameters are precisely known. In this approach, a dedicated probe state and an optimal measurement scheme are designed for each parameter separately, disregarding uncertainties related to the remaining parameters. In contrast, the simultaneous estimation strategy aims to estimate multiple parameters at once using a single quantum probe state and a unified measurement protocol. This approach fully leverages the multiparameter structure of the QFIM, potentially enabling enhanced metrological precision through the use of shared quantum resources.
To explicitly compare the performance of these two strategies, we evaluate the ratio of the minimal total variances associated with parameter estimation in the individual and simultaneous schemes, defined as \cite{Nichols2018,Yousefjani2017,Abouelkhir2023}
\begin{equation}
    \Gamma = \frac{\delta_{ind}}{\delta_{sim}},\label{Gama-G}
\end{equation}
where $\delta_{ind}=\sum_{\varphi_\mu} \mathrm{F}^{-1}_{\varphi_\mu\varphi_\mu}$, and $\delta_{sim}=k^{-1}\;\text{Tr}[\mathrm{F}^{-1}]$, 
here $k$ refers to the number of unknown parameters to be estimated, a factor introduced to account for the resource reduction inherent in the simultaneous estimation strategy discussed above. In general, the ratio $\Gamma\leq k$. A value of $\Gamma>1$ indicates that simultaneous estimation offers an advantage over the individual estimation scheme.
In our case, since two parameters are to be estimated, we set $k=2$. Accordingly, the two quantities $\delta_{ind}$ and $\delta_{sim}$ become
\begin{align}       
\delta_{ind}=&\mathrm{F}^{-1}_{\varphi_\mu\varphi_\mu}+\mathrm{F}^{-1}_{\varphi_\nu\varphi_\nu}, \\  \delta_{sim}=&\frac{1}{2\,\text{det}(\mathrm{F})}\left(\mathrm{F}_{\varphi_\mu\varphi_\mu}+\mathrm{F}_{\varphi_\nu\varphi_\nu}\right).
\end{align}
By substituting these expressions into Eq. \eqref{Gama-G},  we obtain the following form for the ratio $\Gamma$
\begin{equation}
    \Gamma = \frac{2\;\text{det}(\mathrm{F})}{\mathrm{F}_{\varphi_\mu\varphi_\mu}.\mathrm{F}_{\varphi_\nu\varphi_\nu}},
\end{equation}
with $\text{det}(\mathrm{F})=\mathrm{F}_{\varphi_\mu\varphi_\mu}.\mathrm{F}_{\varphi_\nu\varphi_\nu}-\mathrm{F}_{\varphi_\mu\varphi_\nu}.\mathrm{F}_{\varphi_\nu\varphi_\mu}$.

\subsection{ Classical Context}
We consider an efficient measurement strategy based on positive operator-valued measures (POVMs), denoted $\{\hat{\Pi}_x\}$, belonging to the set of all admissible quantum measurements. This method aims to minimize the mean squared error (MSE) in the estimation of unknown parameters, while also allowing for better saturation of the QCRB in the case of an unbiased estimator. Each measurement operator $\hat{\Pi}_x$ satisfies the completeness relation; $\sum_x \hat{\Pi}_x^\dagger \hat{\Pi}_x =\mathbb{1} $,
as required for a consistent description of a quantum measurement process \cite{Degen2017,Giovannetti2011,Giovannetti2006,Liu2020}. 
Once the measurement is fixed, the quantum system yields a conditional probability distribution $\mathcal{P}(x/\varphi)$, which depends on the unknown parameter $x$. This distribution allows the definition of the classical Fisher information (CFI), which measures how sensitively the probability distribution responds to changes in the parameter of interest. The corresponding expression for this probability is provided below
\begin{equation}
    \mathcal{P}(x/\varphi) = Tr[\rho_{\varphi}\Pi_x],
\end{equation}
this conditional probability describes the outcome distribution for estimating the parameter $x$ after the measurement process, where $\varphi$ represents the true value of the parameter. Consequently, to assess the extent to which a measurement outcome reveals information about the parameter, the CFI has been introduced in Refs.~\cite{Degen2017,Giovannetti2011,Giovannetti2006,Liu2020} and can be expressed as
\begin{equation}
    \mathcal{F}^M_\varphi = \int \frac{1}{\mathcal{P}(x/\varphi)} \left[\frac{\partial \mathcal{P}(x/\varphi)}{\partial\varphi}\right]^2dx.\label{CFI Ge}
\end{equation}

To achieve the equality $\mathcal{F}^M_\varphi=\mathrm{F}_\varphi$ ($\mathcal{F}^M_\varphi\leq\mathrm{F}_\varphi$) that is, to make the CFI coincide with the QFI — it is essential to adopt an optimal measurement strategy. In the case of single-parameter estimation, such an optimal strategy can be constructed based on the eigenbasis of the SLD operator $\hat{\mathcal{L}}_\varphi$, defined by the equation $2\partial_\varphi\rho_\varphi = \hat{\rho}_\varphi \,\hat{\mathcal{L}}_\varphi + \hat{\mathcal{L}}_\varphi \,\hat{\rho}_\varphi$. However, implementing this optimal measurement in practice is often highly demanding and may even be infeasible under realistic experimental conditions. This implies that the maximum estimation precision predicted by the QFI may not always be practically attainable.\\
In this work, we focus on evaluating the CFI derived from heterodyne detection. This approach is motivated by the observation that, in certain scenarios, the CFI obtained via heterodyne measurement can closely approach the enhanced QFI provided by our proposed scheme. Heterodyne detection is a widely used Gaussian measurement technique, where the signal field is combined with a reference (or probe) field oscillating at a distinct frequency. The measurement process is formally described using a set of projection operators based on coherent states, namely $\left\{|\alpha\rangle\langle\alpha|/\pi\right\}$ \cite{Aspelmeyer2014}. Based on this formalism, the corresponding expression for the CFI under heterodyne detection can be derived, as detailed in references \cite{Monras2013,Zhang2019,Peng2025}
\begin{equation}
    \mathcal{F}_\varphi = \frac{1}{2} Tr\left[\left(\sigma^{-1} \,\partial_\varphi \sigma\right)^2\right] + \partial_{\varphi}  d_r^{\mathbf{T}} \,\sigma^{-1}\, \partial_{\varphi}  d_r. \label{F clasic}
\end{equation}
The symbol “Tr” refers to the trace operator applied to a matrix. The updated covariance matrix is defined as
$\sigma=\mathcal{V}+\mathbb{1}_6$, where $\mathcal{V}$ characterizes the studied system and “$\mathbb{1}_6$” is the $6\times6$ identity matrix.

\section{APPLICATION} \label{sec_4}
This section explores how different physical parameters affect the precision with which the magnomechanical coupling strength $G_{ms}$ and the magnon–cavity coupling $G_{mc}$ can be estimated. Particular attention is given to the role of the Fisher information in enhancing the accuracy of parameter estimation for both couplings. We perform a comparative analysis between the CFI and the QFI to elucidate their respective capabilities in characterizing the estimation performance. Furthermore, we identify the most informative QCRB, denoted by $\mathcal{B}_{MI}$, which sets the ultimate precision limit for the simultaneous estimation of $G_{ms}$ and  $G_{mc}$. To this end, based on Eqs. \eqref{F with inv}, \eqref{H with inv}, and \eqref{F clasic}, we compute the corresponding QFIMs and the CFI as follows

\begin{align}
     \mathrm{F}_{G_{mi}G_{mj}} =& \,2\, \text{vec}\left[\partial_{G_{mi}}\mathcal{V} \right]^\dagger  
     \mathrm{\Sigma}^{-1}  \text{vec}\left[\partial_{G_{mj}}\mathcal{V} \right]\nonumber \\
     &+ 2 \partial_{G_{mi}}  d_r^{\mathbf{T}} \mathcal{J}^{-1} \partial_{G_{mj}}  d_r \, ,\label{F fin}
\end{align}
\begin{align}
     \mathrm{H}_{G_{mi}G_{mj}} = &\, 2\,\text{vec}\left[\partial_{G_{mi}}\mathcal{V} \right]^\dagger  
     \mathfrak{M}^{-1}  \text{vec}\left[\partial_{G_{mj}}\mathcal{V} \right] \nonumber\\
     &+  \partial_{G_{mi}}  d_r^{\mathbf{T}} \mathcal{V}^{-1} \partial_{G_{mj}}  d_r\,,\label{H fin}
\end{align}
\begin{equation}
    \mathcal{F}_{G_{mj}} = \frac{1}{2} Tr\left[\left(\sigma^{-1} \,\partial_{G_{mj}} \sigma\right)^2\right] + \partial_{G_{mj}}  d_r^{\mathbf{T}} \,\sigma^{-1}\, \partial_{G_{mj}}  d_r\,, \label{F fin}
\end{equation}
where $i$ and $j$ refers either to the optical mode (denoted by $c$) or the mechanical mode (denoted by $s$), and $d_r$ represents the displacement vector of the system. We now proceed with numerical calculations due to the analytical complexity of the expressions derived in the previous equations for the studied system. To identify the most effective method for optimizing the estimation of the coupling strengths $G_{mc}$ and $G_{mb}$, we evaluate the ratio $\mathcal{R}$, as defined in Eq. \eqref{R genarale}. The  experimentally feasible parameters considered in this analysis are as follows\cite{Zhang2016,Rameshti2022,Li2018,Zuo2024}; The cavity, magnon, and mechanical mode frequencies are set to $\omega_c/2\pi=\omega_m/2\pi=10\;\text{GHz}$, $\omega_s/2\pi=10\;\text{MHz}$, with corresponding dissipation rates of $\kappa_c/2\pi=6\times\kappa_m/2\pi=6\;\text{MHz}$, $\gamma_s/2\pi=100\;\text{Hz}$ . The coupling strengths are chosen as $G_{mc}/2\pi=G_{ms}/2\pi=6\;\text{MHz}$, and the analysis is carried out at low temperature $T=50\;\text{mK}$. The cavity and magnon modes are both detuned by an amount equal to the mechanical resonance frequency, i.e., $\Delta_c = \Delta_m = \omega_s$.\par

\begin{figure}[H]
    \centering
    \includegraphics[width=0.45\textwidth]{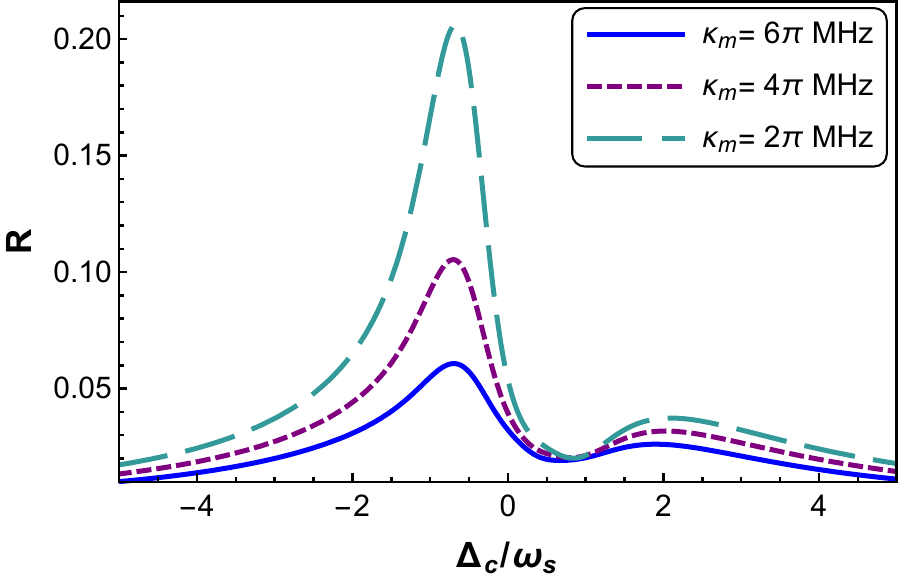}
    \caption{Ratio between the QCRB based on RLD and that based on SLD as a function of the cavity mode detuning $\Delta_c$, under varying dissipation rates $\kappa_m$ of the magnon mode.} \label{R}
\end{figure}

Fig.\ref{R} shows the evolution of the ratio between the two QCRB-RLD and QCRB-SLD as a function of the cavity mode detuning, for different values of the magnon mode dissipation rate $\kappa_m$. It is observed that the ratio $\mathcal{R}$ vanishes when 
$\Delta_c/\omega_s<-5$ and $\Delta_c/\omega_s>5$, and increases as $\kappa_m$ decreases. The ratio reaches its maximum at $\Delta_c \approx -\omega_s$, then decreases as $\Delta_c$ increases. These results indicate that $\mathcal{R}$ remains strictly less than 1 for all values of the detuning $\Delta_c$ and the dissipation $\kappa_m$.
More specifically, the bound $\mathcal{B}_{MI}$ coincides with the bound $\mathcal{B}_R$ for various values of the system parameters.
Overall, the QCRB based on the RLD yields better estimation performance for the systems coupling parameters compared to that based on the SLD. Therefore, the estimation is more accurate for different parameter configurations. This plot was selected solely for the purpose of presenting a concise illustration.

\begin{figure}[H]
    \centering

    % Première image
    \includegraphics[width=0.9\linewidth]{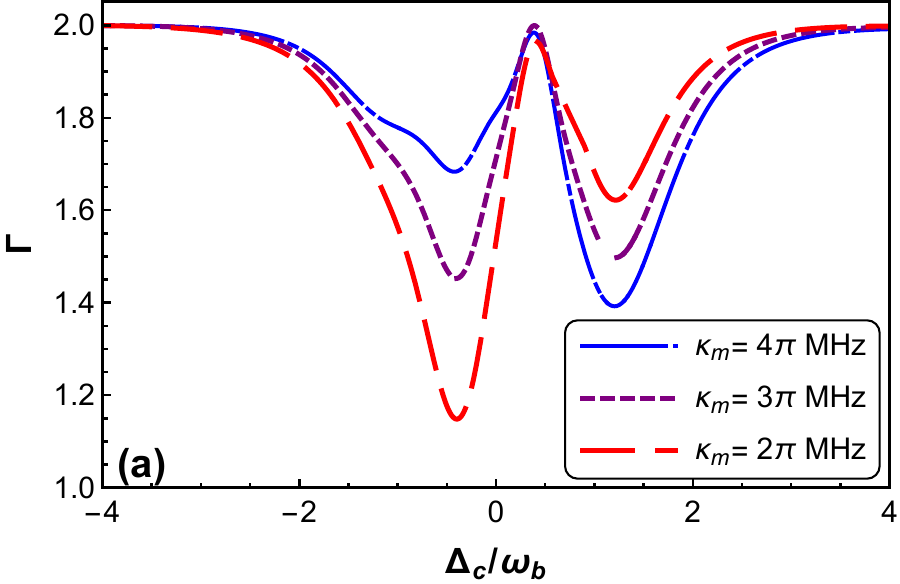}
    %\vspace{0.1cm}
    
    % Deuxième image
    \includegraphics[width=0.9\linewidth]{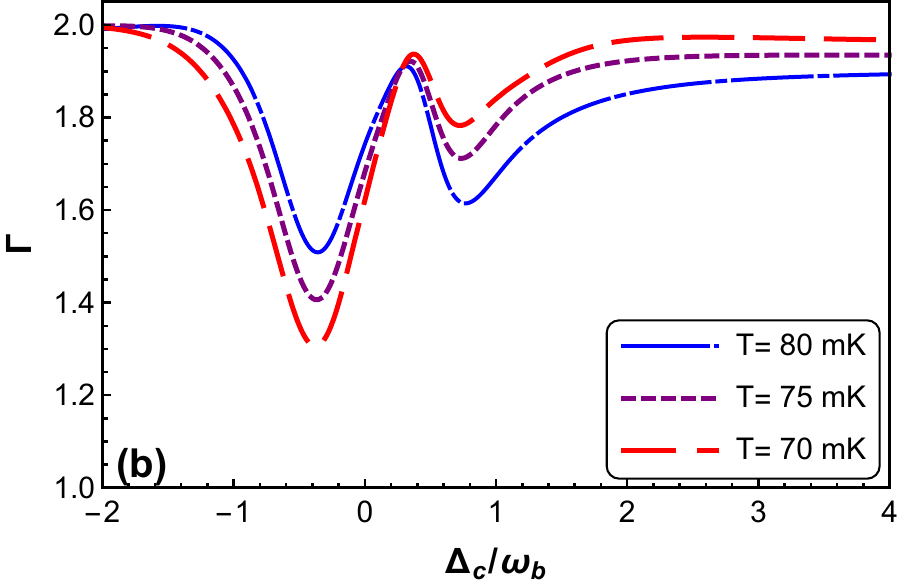}
    %\vspace{0.1cm}

    \caption{Dynamics of the performance ratio $\Gamma$ for the estimation of the constant couplings $G_{mc}$ and $G_{ms}$ as a function of the cavity detuning $\Delta_c$, for various values of magnon dissipation rate $\kappa_m$ in (a), and for different temperatures $T$ in (b). In panel (a), the temperature is fixed at $T=60 \,\text{mK}$, and all other parameters are identical to those used in Fig. \ref{R}.}
    \label{SIM and IND}
\end{figure}

To assess the effectiveness of different metrological strategies, we compare the simultaneous and individual estimation approaches for the two coupling strengths, $G_{mc}$ (magnon–cavity) and $G_{ms}$  (magnon–mechanical), as functions of the cavity detuning $\Delta_c$ , under varying conditions of \ref{SIM and IND}-(a) magnon dissipation rate $\kappa_m$ and \ref{SIM and IND}-(b) environmental temperature $T$. As illustrated in the corresponding figures, the performance ratio $\Gamma$ — defined as the ratio between the precisions achieved via simultaneous and individual estimation — reaches its highest values in the regimes 
$\Delta_c/\omega_s\leq-2$ and $\Delta_c/\omega_s\geq3$, clearly indicating that no advantage is gained through independent estimation of the coupling parameters in these regions. Furthermore, the behavior of $\Gamma$ reveals that it increases with rising $\kappa_m$ and temperature when $\Delta_c/\omega_s<0.5$, while it improves with decreasing $\kappa_m$ and $T$ beyond $\Delta_c/\omega_s > 0.5$. This trend highlights the presence of a locally optimal regime near  $\Delta_c/\omega_s\approx0.5$, where $\Gamma$ typically peaks around 2. Rather than reflecting a symmetric behavior, this point marks a transition zone where the competing effects of dissipation and thermal noise shift the balance of metrological performance. From a physical perspective, the superiority of the simultaneous strategy in specific regions of detuning stems from enhanced coherence among the optical, magnonic, and mechanical modes, which facilitates a more correlated encoding of information about the two couplings. In this regime, the simultaneous estimation framework takes full advantage of the shared structure of quantum correlations, leading to improved precision that would otherwise be inaccessible through separate estimations.\par

\begin{figure}[H]
    \centering

    % Première image
    \includegraphics[width=0.9\linewidth]{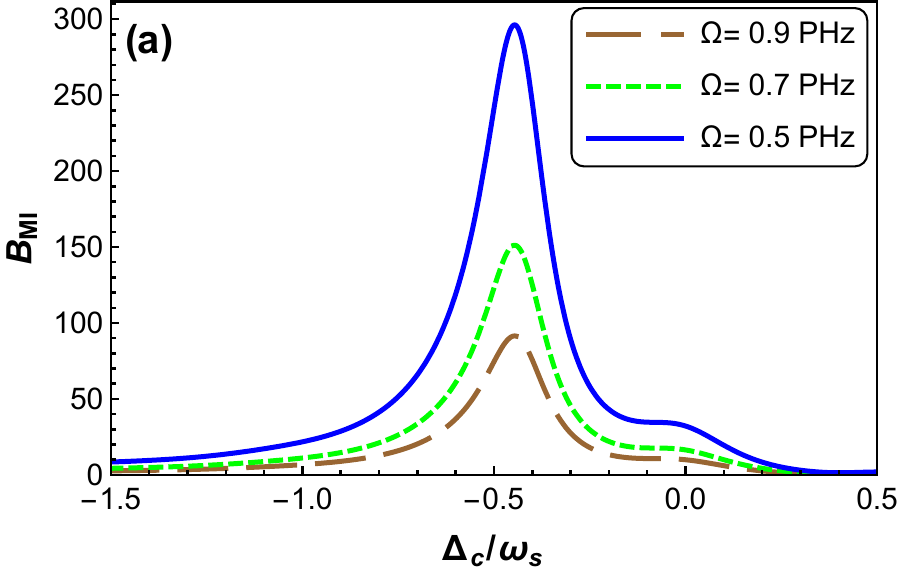}
    %\vspace{0.1cm}
    
    % Deuxième image
    \includegraphics[width=0.9\linewidth]{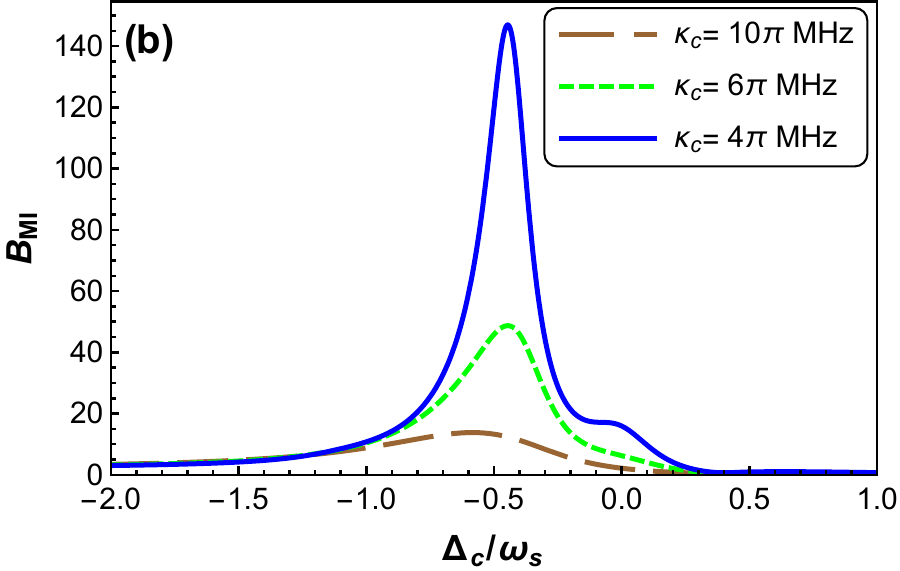}
    %\vspace{0.1cm}
    
    % Deuxième image
    \includegraphics[width=0.9\linewidth]{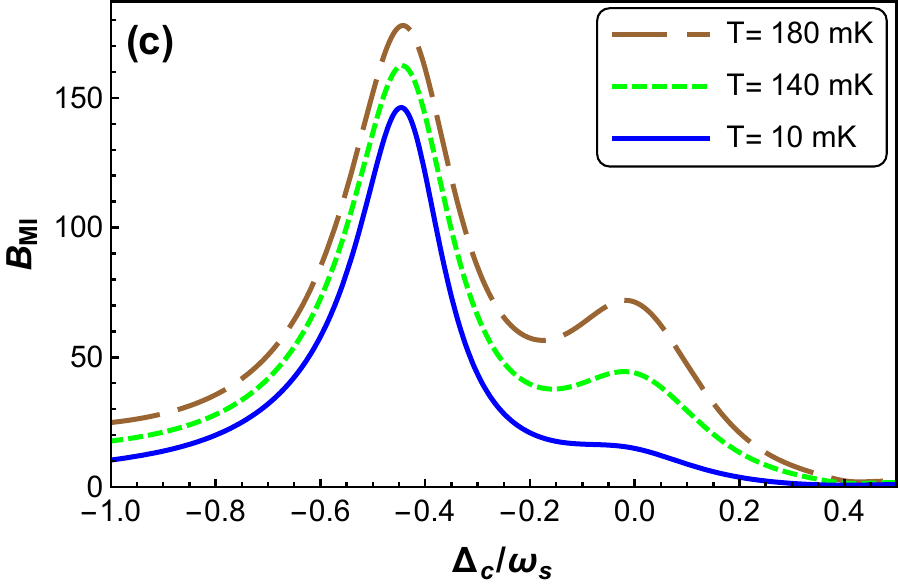}

    \caption{The most informative QCRB ($\mathcal{B}_{MI}$) as a function of the cavity mode detuning $\Delta_c$: (a) under various values of the Rabi frequency $\Omega$, (b) for different cavity mode dissipation rates $\kappa_c$, and (c) for different temperatures $T$. The parameters used are identical to those in Fig. \ref{R}.}
    \label{3 BMI}
\end{figure}

To analyze the parameter estimation precision in a hybrid magnon-mechanical system coupled to an optical cavity, we present in Fig.\ref{3 BMI} the most informative QCRB ($\mathcal{B}_{MI}$), obtained from the QFIM based on the RLD, for the simultaneous estimation of the two fundamental coupling strengths: the magnon-cavity coupling $G_{mc}$ and the magnon-mechanical coupling $G_{ms}$. The QCRB is plotted as a function of the cavity detuning $\Delta_c$ , for \;\ref{3 BMI}-(a) different values of the Rabi frequency $\Omega$, \ref{3 BMI}-(b)\;the cavity dissipation rate $\kappa_c$, and \;\ref{3 BMI}-(c)\;the mechanical mode temperature $T$. The results show that the QCRB exhibits a pronounced maximum around $\Delta_c/\omega_s \approx - 0.5$, indicating an effective resonance between the subsystems, where constructive interference effects reduce the accessibility of quantum information about the parameters, thereby degrading estimation precision. In contrast, a significant improvement in precision (i.e., lower QCRB) is observed when the cavity detuning deviates from this resonant region, particularly for $\Delta_c \leq -\omega_s$ or $\Delta_c/ \omega_s \ge 0.5$, suggesting that such off-resonant regimes allow for more accurate simultaneous estimation of $G_{mc}$ and $G_{ms}$. The parametric analysis also reveals that the QCRB increases as the Rabi frequency decreases, the cavity dissipation rate becomes too low, or the temperature increases. A high Rabi frequency enhances the coherent interactions between the modes, improving the systems sensitivity to parameter variations. Similarly, a moderate cavity dissipation enables an optimal trade-off between information confinement and accessibility, while elevated temperatures introduce thermal noise that degrades the quantum correlations necessary for efficient estimation. These results highlight the importance of precise control over the systems physical parameters in order to maximize accessible information and optimize metrological performance in hybrid quantum platforms.
\begin{figure}[H]
    \centering

    % Première image
    \includegraphics[width=0.9\linewidth]{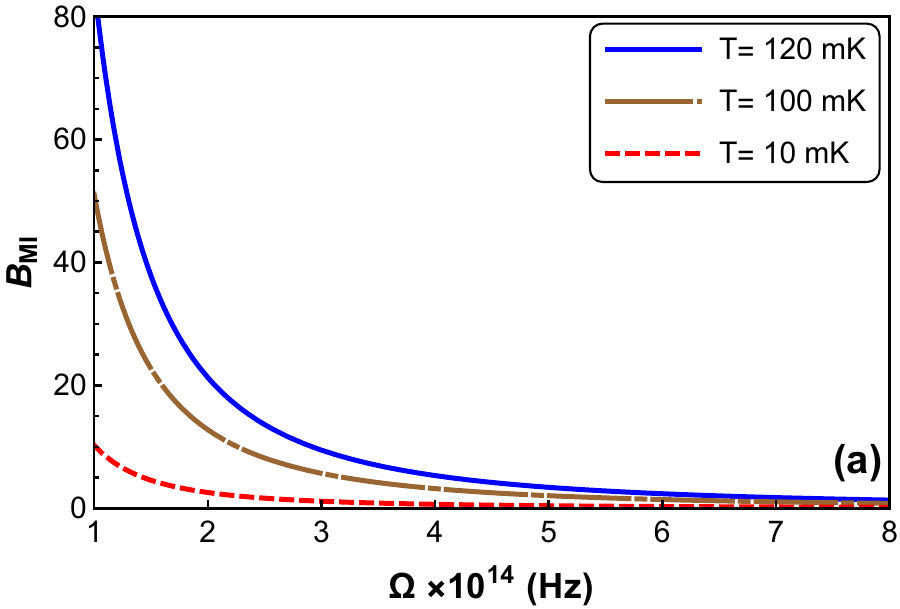}
    %\vspace{0.1cm}
    
    % Deuxième image
    \includegraphics[width=0.9\linewidth]{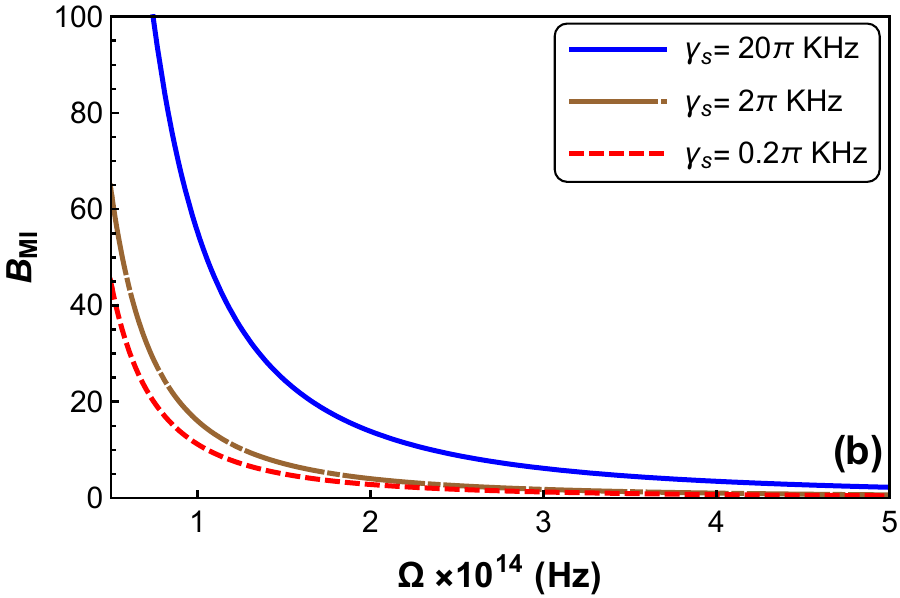}
    %\vspace{0.1cm}
    \caption{Behavior of the most informative QCRB ($\mathcal{B}_{MI}$) as a function of the Rabi frequency $\Omega$: (a) for various temperatures $T$, and (b) under various values of the mechanical damping rate $\gamma_s$. The cavity dissipation rate is fixed at $\kappa_c = 6 \;\text{MHz}$, with all remaining parameters matching those employed in Fig.~\ref{R}.}\label{2BMI}
\end{figure}

Fig.\ref{2BMI} shows the behavior of the QCRB-$\mathcal{B}_{MI}$ with respect to the Rabi frequency $\Omega$, under varying temperature values $T$ Fig.\ref{2BMI} - (a) and mechanical damping rate $\gamma_s$ Fig. \ref{2BMI} - (b), in the context of jointly estimating the magnon-cavity $G_{mc}$ and magnon-mechanical $G_{ms}$ coupling strengths. The results reveal a rapid decrease in the QCRB-$\mathcal{B}_{MI}$ as $\Omega$ increases, indicating a significant improvement in the precision of the parameter estimation. This reduction is attributed to the enhancement of coherent interactions among the subsystems, which increases the systems sensitivity to variations in the target parameters and facilitates the accumulation of useful quantum information. However, this enhancement is considerably hindered when either the temperature $T$ or the damping rate $\gamma_s$ increases: for the same value of $\Omega$, the QCRB-$\mathcal{B}_{MI}$ becomes larger at higher values of $T$ or $\gamma_s$. The effect of $\gamma_s$ is particularly critical, as strong mechanical damping suppresses the mechanical modes oscillations, thereby weakening the correlations among the subsystems and degrading the transmission of information. Consequently, the optimal metrological performance—i.e., more precise estimation of both couplings $G_{mc}$ and $G_{ms}$ —is achieved in regimes characterized by moderate Rabi frequencies, low mechanical temperatures, and reduced mechanical damping. These conditions simultaneously promote quantum information accumulation and preserve the correlations required for efficient joint parameter estimation in the hybrid system.

\begin{widetext}

\begin{figure}[t]
  \centering
  \begin{subfigure}[b]{0.245\textwidth}
    \centering
    \includegraphics[width=\textwidth]{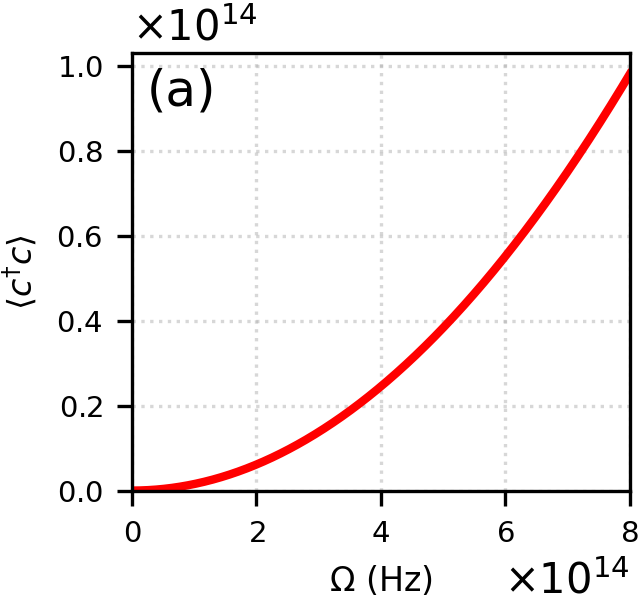}
  \end{subfigure}
  \begin{subfigure}[b]{0.245\textwidth}
    \centering
    \includegraphics[width=\textwidth]{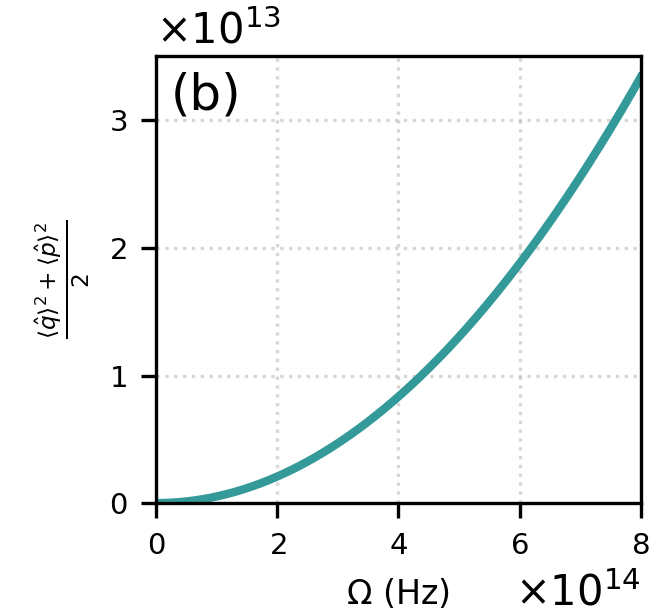}
  \end{subfigure}
  \begin{subfigure}[b]{0.245\textwidth}
    \centering
    \includegraphics[width=\textwidth]{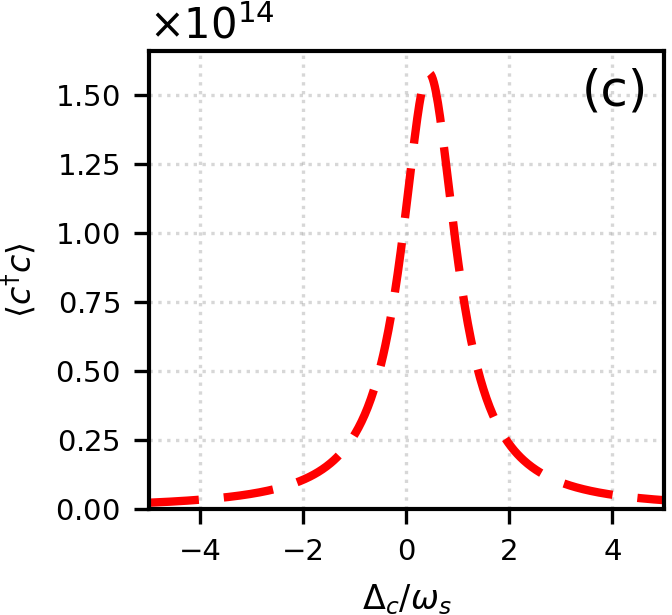}
  \end{subfigure}
  \begin{subfigure}[b]{0.238\textwidth}
    \centering
    \includegraphics[width=\textwidth]{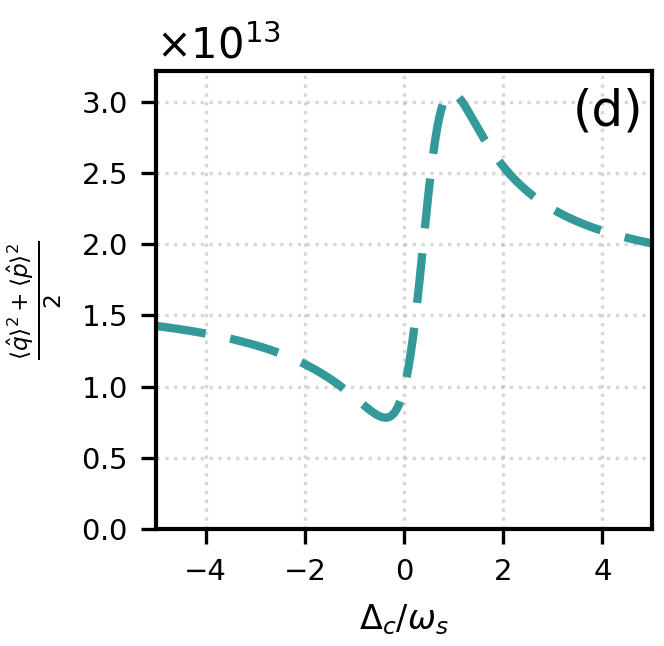}
  \end{subfigure}

  \caption{Plot of the average number of photons[phonons] as a function of(a)[(b)] the driving frequency $\Omega$, and (c)[(d)] the cavity detuning $\Delta_c$. The parameters used are identical to those in Fig. \ref{R}.}
  \label{phot and phon}
\end{figure}

\end{widetext}

\begin{figure}[H]
    \centering
    % Première image
    \includegraphics[width=0.9\linewidth]{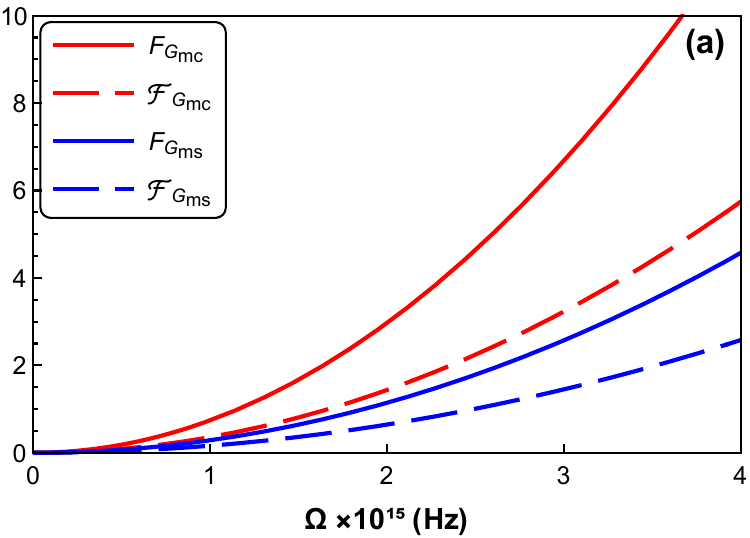}
    % Deuxième image
    \includegraphics[width=0.9\linewidth]{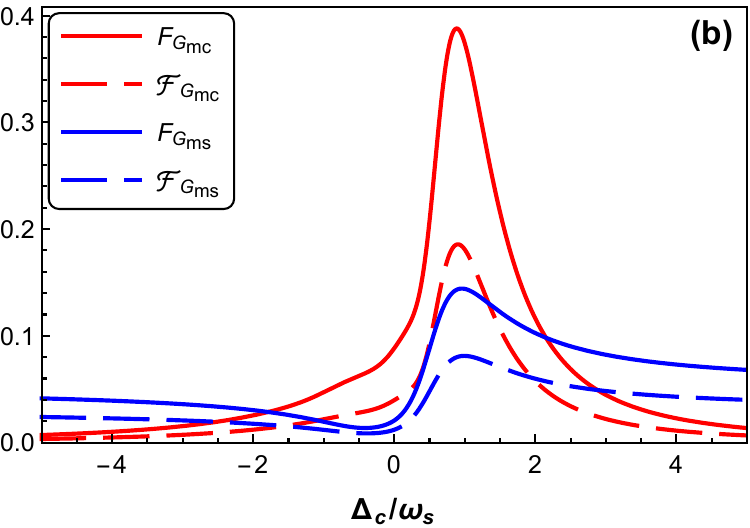}\vspace{-0.15cm}
    \caption{Plot of the QFI (solid lines) and the CFI obtained via heterodyne detection (dashed lines) as a function of: (a) the Rabi frequency $\Omega$, and (b) the cavity detuning $\Delta_c$. Red curves correspond to the estimation of the magnon–cavity coupling $G_{mc}$, while blue curves correspond to the estimation of the magnon–mechanical coupling $G_{ms}$. The parameters used are identical to those in Fig. \ref{R}.}\label{QFI CFI}
\end{figure}

In Fig.\ref{phot and phon}, we display the average numbers of photons and phonons as functions of the Rabi frequency $\Omega$ and the cavity detuning $\Delta_c$. As shown in Figs. \ref{phot and phon}-(a) and \ref{phot and phon}-(b), the average photon and phonon numbers increase significantly with increasing Rabi frequency. This behavior serves as a key mechanism through which the driving field enhances the precision of parameter estimation. Specifically, a higher average number of photons and phonons in the hybrid cavity-magnon-mechanical system strengthens both the magnon-cavity and magnon-mechanical interactions and enlarges the information storage capacity. Consequently, more information becomes available for estimating the coupling parameters $G_{mc}$ and $G_{ms}$, as evidenced in Fig. \ref{2BMI}, thereby leading to improved estimation precision. Moreover, the dependence of the average photon and phonon numbers on $\Omega$ closely mirrors that of the QFIs $\mathcal{F}_{G_{mc}}$ and $\mathcal{F}_{G_{ms}}$ , as illustrated in Fig. \ref{QFI CFI}-(a), indicating that the Rabi frequency positively influences the estimation precision of the coupling strengths. It is also noteworthy, as shown in Figs. \ref{phot and phon}-(c) and \ref{phot and phon}-(d), that the dependence of the average photon and phonon numbers on $\Delta_c$ exhibits a striking resemblance to the behavior of the QFIs $\mathcal{F}_{G_{mc}}$ and $\mathcal{F}_{G_{ms}}$ with respect to $\Delta_c$, shown in Fig. \ref{QFI CFI}-(b). At a specific value of $\Delta_c$, both photon and phonon numbers reach a maximum that coincides with the peak values of the QFI, indicating an optimal interaction point. This simultaneous maximization of intracavity and mechanical excitations, along with the QFIs, reflects a heightened sensitivity of the system to variations in the coupling parameters, enabling more precise estimation. These optical and mechanical excitations thus represent essential resources for optimizing the metrological performance of the system. This highlights the pivotal role of the cavity detuning $\Delta_c$, which simultaneously regulates the accumulation of information about the couplings by modulating the effective resonance between the modes, thereby serving as a critical tuning parameter for enhancing metrological precision.

Fig.\ref{QFI CFI} displays the Fisher information curves corresponding to the separate estimation of the two coupling parameters in the hybrid cavity–magnon–mechanical system: the magnon–cavity coupling $G_{mc}$ (in red) and the magnon–mechanical coupling $G_{ms}$ (in Blue). These quantities are plotted as functions of the Rabi frequency $\Omega$ and the cavity detuning $\Delta_c$, with solid lines representing the QFI calculated using the RLD formalism, and dashed lines corresponding to the CFI obtained via heterodyne detection. As shown in Fig. \ref{QFI CFI}-(a), the Fisher information is relatively low for small values of $\Omega$, but increases significantly as $\Omega$ grows. Physically, this behavior stems from the fact that a higher Rabi frequency enhances the amplitude of the driving field injected into the cavity, thereby increasing the intracavity photon number. This excitation propagates through the magnon–cavity and magnon–mechanical interactions, stimulating both magnetic and mechanical modes. As a result, the coupling parameters $G_{mc}$ and $G_{ms}$ exert a stronger influence on the global state, making the system more sensitive to their variations. Consequently, the state (or the measurement probability distribution) exhibits a steeper derivative with respect to these couplings, leading to an increase in Fisher information, which reflects the systems enhanced ability to encode and extract information about the unknown parameters. Furthermore, Fig. \ref{QFI CFI}-(b) reveals that the Fisher information reaches a maximum when $\Delta_c \approx\omega_s$, the mechanical frequency, indicating an effective resonance between the three modes of the system. In this tripartite resonance regime, energy and information are transferred coherently and efficiently between subsystems, amplifying the effect of both couplings and enhancing parameter sensitivity. This resonance condition results in a peak of both QFI and CFI, signaling an optimal regime for parameter estimation. Comparing the CFI obtained from heterodyne detection $\mathcal{F}_{G_j}$ with the QFI $\mathrm{F}_{G_j}$ , we observe that the inequality $\mathcal{F}_{G_j} \leq \mathrm{F}_{G_j} $, $(j=mc, ms)$ is always satisfied, and the two quantities are very close across the parameter range. This suggests that heterodyne detection, which exploits the frequency shift between two optical fields to measure the quadratures associated with position and momentum, allows for near-optimal information extraction. In the Gaussian regime, where all relevant information is encoded in the first-order moments and the covariance matrix, analyzing the interference signal enables the reconstruction of a probability distribution that is highly sensitive to the coupling parameters. From this distribution, and using Eq. \eqref{CFI Ge}, a CFI can be extracted that closely approximates the QFI. Although the optimal measurement corresponding to the SLD can be theoretically constructed from the moments and the covariance matrix, its experimental implementation remains challenging. Finally, it is worth noting that the Fisher information associated with $G_{mc}$ is consistently higher than that of $G_{ms}$, regardless of the value of $\Omega$ or $\Delta_c$. This asymmetry arises from the systems structure: $G_{mc}$ directly links the driven optical field to the magnons, making its estimation more sensitive to changes in control parameters. In contrast, $G_{ms}$, which only acts indirectly through the magnons, has a more limited effect on measurable observables, resulting in a lower estimation precision.

\section{Conclusion} \label{sec_5}
In hybrid quantum systems, quantum metrology aims to precisely estimate the magnon–photon and magnon–phonon coupling strengths, which is essential for understanding, controlling, and optimizing the performance of cavity magnomechanical devices. The coupling $G_{mc}$, arising from the magnetic interaction between magnons and the microwave field of a cavity, enables coherent excitation exchange and leads to characteristic phenomena such as polariton formation and vacuum Rabi splitting. Meanwhile, the coupling $G_{ms}$, stemming from the magnetostrictive effect, mediates the interaction between magnons and mechanical modes of the system. Combined, these interactions allow for the harnessing of tripartite photon–magnon–phonon dynamics in a variety of quantum technologies, including information processing and high-precision sensing.\\
In the present work, we investigated a cavity magnomechanical system by analyzing the fundamental limits of the simultaneous estimation of both couplings through the QCRB. We explicitly computed the QFIMs associated with the RLD and the SLD operators within the framework of multimode Gaussian states. A comparison between simultaneous and individual estimation strategies revealed a clear advantage for the simultaneous approach, demonstrating its relevance for multiparameter quantum metrology. We also derived the saturation condition of the QCRB under the RLD formalism and applied it to a Fabry–Pérot cavity system, thereby highlighting the theoretical scope of the method. Furthermore, the CFI obtained from heterodyne detection was compared to the QFI, suggesting that this measurement approach provides an effective strategy for parameter estimation.\\
Our analysis showed that the precision of estimating the two coupling strengths decreases as the cavity detuning approaches $-\omega_s/2$. Conversely, increasing the Rabi frequency and the cavity dissipation rate, along with reducing mechanical damping and temperature, significantly lowers the QCRB, indicating an improvement in the amount of information accessible about the estimated parameters. We also found that the behavior of the average photon and phonon numbers mirrors that of the Fisher information: increasing these numbers further enhances estimation precision by reducing the QCRB.\\
These findings further demonstrate that heterodyne detection provides an effective strategy for jointly estimating the coupling strengths in cavity magnomechanical systems, thereby opening up promising avenues for the advancement of hybrid quantum metrology protocols.
\section*{Declarations} 

{\bf Data Availability Statement:} No data were used for the research described in the article.\\

{\bf Conflict of interest:} The authors declare that they have no known competing financial interests or personal relationships that could have appeared to influence the work reported in this paper.

\end{document}